\pdfoutput=1
\documentclass[namedreferences,hyperref,optionalrh]{spr-sola}
\usepackage{graphicx}        
\usepackage{amsmath}
\usepackage{amssymb}
\usepackage{color}           

\newcommand{\aap}{{\it Astron. Astrophys.}}

\newcommand{\apj}{{\it Astrophys. J.}}
\newcommand{\apjl}{{\it Astrophys. J. Lett.}}

\newcommand{\mnras}{{\it Mon. Not. R. Astron. Soc.}}

\newcommand{\solphys}{{\it Solar Phys.}}
 
\newcommand{\ssr}{{\it Space Sci. Rev.}} 
\chardef\us=`\_

\begin{document}

\begin{frontmatter}
\title{Solar poloidal magnetic field generation rate from observations and
mean-field dynamos}
\author[addressref={aff1},email={valery.pipin@gmail.com}]{\inits{V.V.}\fnm{Valery}~\snm{Pipin}\orcid{0000-0001-9884-1147}}
\address[id=aff1]{Institute solar-terrestrial physics, Irkutsk,
  Russia}

\runningauthor{V.V. Pipin}
\runningtitle{Large-scale poloidal magnetic flux
  generation rate}

\begin{abstract}
To estimate the hemispheric flux generation rate of the large-scale radial magnetic field in the Solar Cycles 23 and 24, we use the photospheric observations of the solar magnetic fields and results of the mean-field dynamo models. Results of the dynamo model show the strong impact of the radial turbulent diffusion on the surface evolution of the large-scale poloidal magnetic field and on the hemispheric magnetic flux generation rate. {To process the observational data set we employ the parameters of the meridional circulation and turbulent diffusion from the Surface Flux-Transport (SFT) models. We find that the observed evolution of the axisymmetric vector potential contains the time--latitude patterns which can result from the effect of turbulent diffusion of the large-scale poloidal magnetic field in the radial direction. We think that, the SFT models can reconcile the observed rate of hemispheric magnetic flux generation by considering radial turbulent diffusion and lower values of the diffusion coefficient.}
\end{abstract}
\end{frontmatter}

\section{Introduction}
The basic scenarios of the solar dynamo cycle base on the large-scale hydromagnetic dynamo mechanism of \citet{Parker1955}. Following Parker’s idea, the large-scale toroidal magnetic field, which is generated in the depth of the solar convection zone from the poloidal magnetic field by the differential rotation, forms the 11-year sunspot cycle. In addition, the large-scale poloidal magnetic field can be regenerated because of effects of the cyclonic convective motions. The dynamo process of the large-scale poloidal magnetic field generation is under debate \citep{CharSok2023SSRv}. There are two basic alternatives: 1) the turbulent $\alpha$ effect, which is caused by the broken reflection symmetric of the turbulent motions \citep{Krause1980}; 2) the so-called Babcock-Leighton effect, which usually relate  with the tilt of the emerging BMRs toward the solar equator  \citep{Leighton1969}. Formally, the Babcock--Leighton effect can be considered as a special case of the mean-field $\alpha$ effect \citep{Hazra2023a}. The alternative dynamo scenarios result in a different location of the dynamo sources of the large-scale poloidal magnetic field in the solar convection zone. When the turbulent $\alpha$ effect dominates, the poloidal magnetic field is generated in the bulk of the convection zone and propagated toward the surface, where it is redistributed by the turbulent diffusion and the meridional circulation. This is the case of the so-called distributed dynamo \citep{Brandenburg2005,BRetal23}. In the Babcock--Leighton scenario, the large-scale poloidal magnetic field is generated in the near-surface layer as a result of the averaged effect of the emerging BMRs \citep{Hazra2023a}. Such models try to take into account the phenomenology of the surface magnetic field evolution on the Sun. It is noteworthy that the predicted time-latitude diagrams of the large-scale radial magnetic field on the surface look very similar for both scenarios. In our recent paper \citep{PKT23}, we showed the emergence of the BMRs on the surface and that their effect on the large-scale dynamo can be taken in the 3D mean-field dynamo model. The given model reproduces quantitatively the basic parameters of the Solar Cycles 23 and 24. Similar to the 3D flux-transport models of  \cite{Miesch2014Ap} and \cite{Kumar2019} our model uses internal toroidal magnetic field as a source of the BMRs formation. However, in our case, the large-scale poloidal magnetic field in the bulk of the convection zone is regenerated by the $\alpha$ effect. The interesting point is that, for the weakly nonlinear case, when the mean-field dynamo operates with  the $\alpha$ effect, which is close to the dynamo threshold, the surface BMRs' activity provides a considerable impact on the dynamo. 

Naturally, the question arises if it is possible to filter out the
basic dynamo scenarios using the solar magnetic field observations
and results of the dynamo models? To answer the question we consider
the flux of the axisymmetric poloidal magnetic field,
$\overline{\boldsymbol{B}}^{(P)}$,
for example, through the northern hemisphere of the Sun, $\Phi^{N}=\int_{N}\overline{\boldsymbol{B}}^{(P)}\cdot\mathrm{d}\boldsymbol{S}$.
Using Stokes's theorem, we transform it in the line integral over the
solar equator, 
\begin{eqnarray}
\Phi^{N} & = & \int_{N}\overline{\boldsymbol{B}}^{(P)}\cdot\mathrm{d}\boldsymbol{S}=\int_{N}\boldsymbol{\nabla}\times\left(\boldsymbol{\hat{\phi}}A\right)\cdot\mathrm{d}\mathbf{S}\label{eq:phin}\\
 & = & \oint_{eq}A\boldsymbol{\hat{\phi}}\cdot\mathrm{d}\boldsymbol{\mathbf{\ell}}=2\pi RA\rvert_{eq},\label{eq:aeq}
\end{eqnarray}
where $\hat{\boldsymbol{\phi}}A$ is the axisymmetric vector potential, $\hat{\boldsymbol{\phi}}$ is the unit vector in the azimuthal direction, and $R$ is the solar radius. Equation \ref{eq:aeq} is more practical than  Equation \ref{eq:phin} for estimation of $\Phi^{N}$ using the data of the dynamo models. In this paper, we employ the results of the dynamo models from our previous paper \citep{PKT23}. The next section describes briefly the dynamo model and the algorithm of computation of the poloidal flux generation rate parameters. Section 3 shows the data reduction. Section 4 shows the main results of our analysis and the last section contains discussion and basic conclusions of our paper.

\section{Dynamo model and poloidal magnetic field generation rate}

We consider the mean-field evolution equation for the induction vector
of the large-scale magnetic field, 
\begin{eqnarray}
\partial_{t}\left\langle \boldsymbol{B}\right\rangle  & = & \mathbf{\nabla}\times\left(\mathbf{\boldsymbol{\mathbf{\mathcal{E}}}}+\mathbf{\boldsymbol{\mathbf{\mathcal{E}}}^{\mathrm{(BMR)}}}+\overline{\boldsymbol{U}}\times\left\langle \boldsymbol{B}\right\rangle \right),\label{eq:dyn}
\end{eqnarray}
where, we decompose the induction vector $\left\langle \mathbf{B}\right\rangle$ into the sum of the
axisymmetric and pure nonaxisymmetric parts: 
\begin{eqnarray}
\left\langle \mathbf{B}\right\rangle  & = & \overline{\mathbf{B}}+\tilde{\mathbf{B}}.\label{eq:b0}
\end{eqnarray}
Hereafter, we denote by $\left\langle \dots\right\rangle $ the mean over ensemble of the turbulent field, and the overbar $\overline{\dots}$ denotes the azimuthal averaging. The turbulent electromotive force, $\mathbf{\mathcal{E}}=\left\langle \boldsymbol{u}\times\boldsymbol{b}\right\rangle $ is determined by the randomly fluctuating turbulent velocity and magnetic field, $\boldsymbol{u}$ and $\boldsymbol{b}$, respectively. Following  \citet{PKT23} we introduce the special part of the mean electromotive force, $\mathbf{\boldsymbol{\mathbf{\mathcal{E}}}^{\mathrm{(BMR)}}}$, to describe the emergence and effects of the BMRs on the large-scale dynamo. The vector $\overline{\boldsymbol{U}}$ defines the axisymmetric large-scale flows, which include the differential rotation and the meridional circulation.

The turbulent mean electromotive force was calculated analytically
using the mean-field magnetohydrodynamics framework of \citet{Krause1980}.
In general case, it reads as 
\begin{equation}
\mathcal{E}_{i}=\left(\alpha_{ij}+\gamma_{ij}\right)\left\langle B\right\rangle _{j}-\eta_{ijk}\nabla_{j}\left\langle B\right\rangle _{k},\label{eq:emf}
\end{equation}
For injections of the BMRs, we use the mean-field like expression:
\begin{equation}
\mathcal{E}_{i}^{(\mathrm{BMR})}=\alpha_{\beta}\delta_{i\phi}\left\langle B\right\rangle _{\phi}+V_{\beta}\left(\hat{\boldsymbol{r}}\times\left\langle \mathbf{B}\right\rangle \right)_{i},\label{eq:ebmr}
\end{equation}
where the first term describes the alpha effect, which is connected with the tilt of the individual BMR (parameter $\alpha_{\beta}$), and the second term describes the BMRs' emergence in the form of the bipolar magnetic region. The NOAA database (https://www.swpc.noaa.gov/) was used to estimate the location, size, emergence time, and rate of emergence. The spatial profile of $V_{\beta}$ is  Gaussian with the center and width, which correspond to location and size of the BMRs. The magnitude of  $V_{\beta}$ is determined by the mean-field effects of the turbulent and magnetic buoyancy \citep{Ruediger1995}. The BMRs alpha effect (associated with the BMRs tilt) is considered as the random value, which is fluctuating relative to the mean value corresponding to the Joy’s law, 
\begin{equation}
\alpha_{\beta}=\left(C_{\alpha\beta}\cos\theta+\xi_{\alpha}\right)V_{\beta}\psi_{\alpha}(\beta).\label{eq:ab}
\end{equation}
The parameter $C_{\alpha\beta}$ determines the average tilt of BMRs for the specified latitude. Following to \citet{PKT23}, $C_{\alpha\beta}=0.5$ results in a good agreement with solar observations. The parameter $\xi_{\alpha}$ controls the random fluctuations of the BMRs tilt. Similarly to \citet{Rempel2005c}, the evolution of $\xi_{\alpha}$ is driven by the third--order system of the differential equations, 
\begin{eqnarray}
\dot{\xi}_{\alpha} & = & -\frac{2}{\tau_{\xi}}\left(\xi_{\alpha}-\xi_{1}\right),\label{xia}\\
\dot{\xi}_{1} & = & -\frac{2}{\tau_{\xi}}\left(\xi_{1}-\xi_{2}\right),\nonumber \\
\dot{\xi}_{2} & = & -\frac{2}{\tau_{\xi}}\left(\xi_{2}-g\sqrt{\frac{2\tau_{\xi}}{\tau_{h}}}\right).\nonumber 
\end{eqnarray}
Here, $g$ is the random number from the Gaussian probability distribution; the code renews its value at each time step of the dynamo model, $\tau_{h}$. The choice of $\tau_{h}$ is dictated  by the numerical stability criteria, and it is sufficiently small in comparison with the other typical time scales of the mean-field dynamo processes. The parameter $\tau_{\xi}$  controls the relaxation time scale of perturbation $\xi_{\alpha}$ . Following the above cited paper, we put $\tau_{\xi}=2$ months. In addition, it is assumed that, $\overline{g}=0$, $\sigma\left(g\right)=C_{\alpha\beta}/2$, i.e., the mean-square magnitude of variations, $\sigma$, corresponds to the mean tilt at the latitude of 30$^{\circ}$.

\begin{figure}
  \includegraphics[width=0.95\columnwidth]{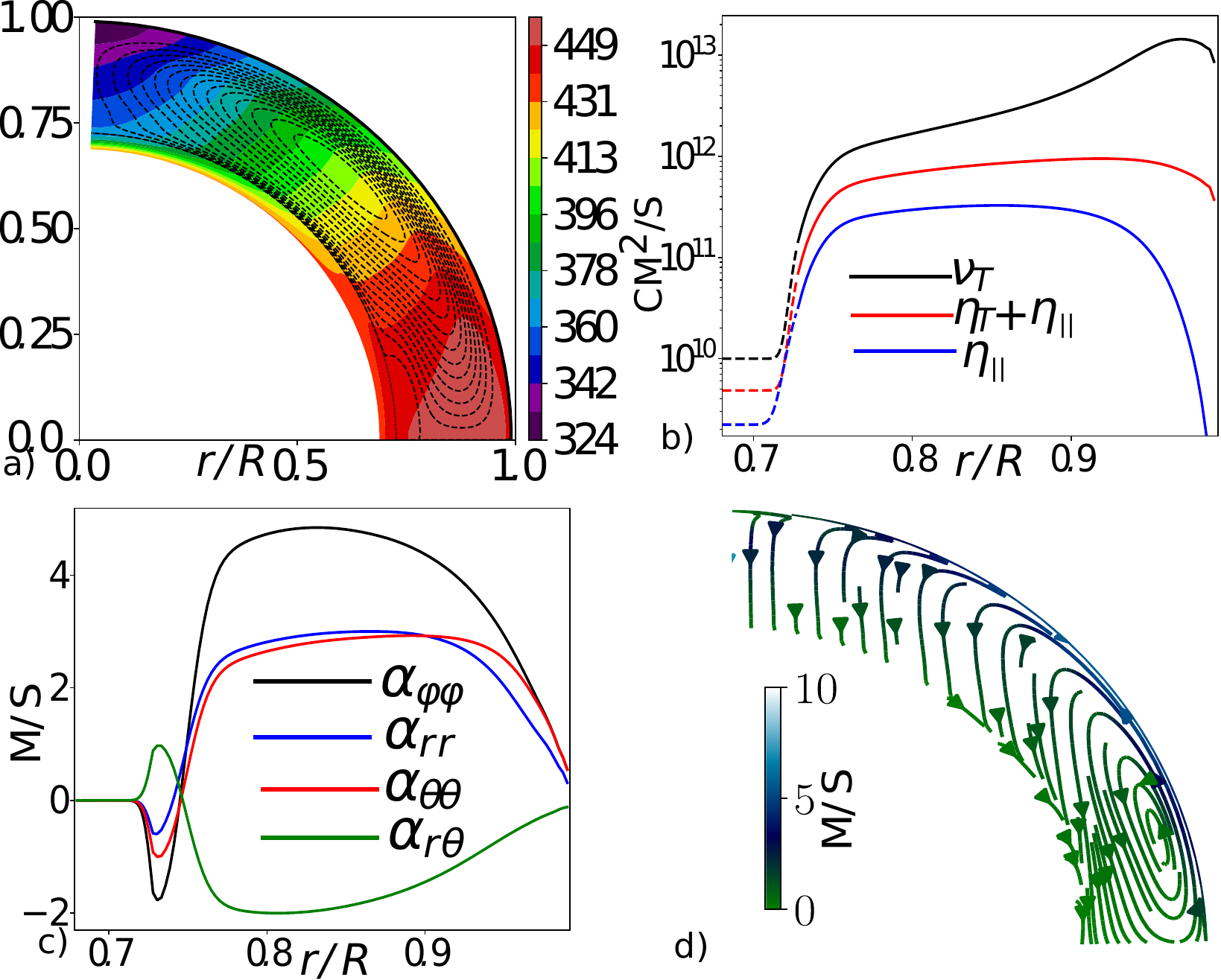}
  \caption{a) The profile of the angular velocity distribution in nHz and the
streamlines of the meridional circulation for the nonmagnetic case
of in the dynamo model, the dashed line corresponds to a counterclockwise
flow; the magnitude of the circulation is 13 m s$^{-1}$ on the surface; b)
radial profiles of the anisotropic eddy diffusivity and the eddy viscosity
at latitude 45$^{\circ}$; c) radial profiles of the alpha-effect
tensor; and d) the streamlines of the effective drift of the toroidal
magnetic field because of the effects of turbulent pumping and the meridional
circulation.}\label{Fig1}
\end{figure}

The papers of \citet{PKT23} and \citet{PK24} describe  $\mathbf{\boldsymbol{\mathbf{\mathcal{E}}}^{\mathrm{(BMR)}}}$in details, including animations of the BMRs emergence and their modeled evolution in the Solar Cycles 23 and 24. The dynamo model calculates the general parameters, including differential rotation and meridional circulation, in a way that incorporates the effect of the generated magnetic field on heat transport and angular momentum balance. These questions are described in more detail in the works cited above. Figure 1 shows the angular velocity distribution profile, the meridional circulation structure, as well as radial profiles of the alpha effect tensor and the turbulent diffusion at latitude 45$^{\circ}$, and the structure of the effective drift of a large-scale magnetic field in the dynamo model. In the upper half of the convective zone, the turbulent diffusion magnitude reaches a value of $10^{12}$ cm$^{2}$s$^{-1}$ and on the surface it is $4\times10^{11}$cm$^{2}$s$^{-1}$. This is approximately consistent with the analysis of observations of the effect of solar supergranulation on the decay of solar active regions \citep{HathChoudh2008SoPh}. Further details of the solar dynamo model under consideration can be found in \citet{PKT23}.

The variation rate of the axisymmetric radial magnetic field flux
in the northern hemisphere is determined by the evolution of the axisymmetric vector potential at the equator,
\begin{eqnarray}
\partial_{t}\Psi_{N} & = & \partial_{t}\int_{N}\mathbf{\overline{B}}_{P}\cdot\mathrm{d}\mathbf{S}=\int_{N}\boldsymbol{\nabla}\times\left(\mathbf{\boldsymbol{\mathcal{E}}+\mathbf{\overline{\boldsymbol{\mathbf{\mathcal{E}}}^{\mathrm{(BMR)}}}}+}\left(\overline{\mathbf{U}}\times\mathbf{\overline{B}}\right)\right)\cdot\mathrm{d}\mathbf{S}\label{eq:a}\\
 & = & 2\pi R\partial_{t}A\left(t,\pi/2,R\right),\nonumber 
\end{eqnarray}
where $\overline{\boldsymbol{\mathbf{\mathcal{E}}}^{\mathrm{(BMR)}}}$
corresponds to the azimuthally averaged effect of BMRs' emergence.
Note that the evolution of the axisymmetric azimuthal vector potential
satisfies the dynamo equation: 
\[
\partial_{t}A=\mathbf{\mathcal{E}_{\phi}}+\overline{\mathcal{E}_{\phi}^{\mathrm{(BMR)}}}+\left(\overline{\mathbf{U}}\times\mathbf{\overline{B}}\right)_{\phi}.\label{eqAev}
\]
Here, the last term reflects the effect of meridional circulation.
The objective of this paper  is to evaluate the relative contribution of the different parts of this equation on evolution of the vector potential. To analyze the various contributions to the evolution,
we categorize the turbulent electromotive force into three individual components, 
\begin{equation}
\mathcal{E}_{\phi}=\mathcal{E_{\phi}^{\alpha}}+\mathcal{E_{\phi}^{\gamma}}+\mathcal{E_{\phi}^{\eta}},\label{eq:ef}
\end{equation}
where $\mathcal{E_{\phi}^{\alpha}}$ corresponds to effects of the
turbulent generation of the large-scale poloidal magnetic field, $\mathcal{E_{\phi}^{\gamma}}$
stands for the turbulent pumping, and $\mathcal{E_{\phi}^{\eta}}$
reflects the effect of the turbulent diffusion. General analytical
expressions for these contributions can be found in \citet{Pipin2008a}
or in the recent paper by \citet{P22}. We distinguish the turbulent diffusion on the radial coordinate. It is denoted as  $\mathcal{E}_{\phi}^{\eta(rad)}$. This contribution is proportional to ${\displaystyle \frac{1}{r}\frac{\partial^{2}rA}{\partial r^{2}}}$ , it is also associated with the radial derivative of the meridional component of the large-scale magnetic field. We will also compute the effect of turbulent diffusion on latitude, which is hereafter denoted as $\mathcal{E}_{\phi}^{\eta(lat)}$ . It is proportional to ${\displaystyle \frac{\sin\theta}{r^{2}}\frac{\partial^{2}A}{\partial\mu^{2}}}$, where $\mu=\cos\theta$ and $\theta$ is the polar angle.

\begin{figure}[ht]
  \includegraphics[width=0.95\columnwidth]{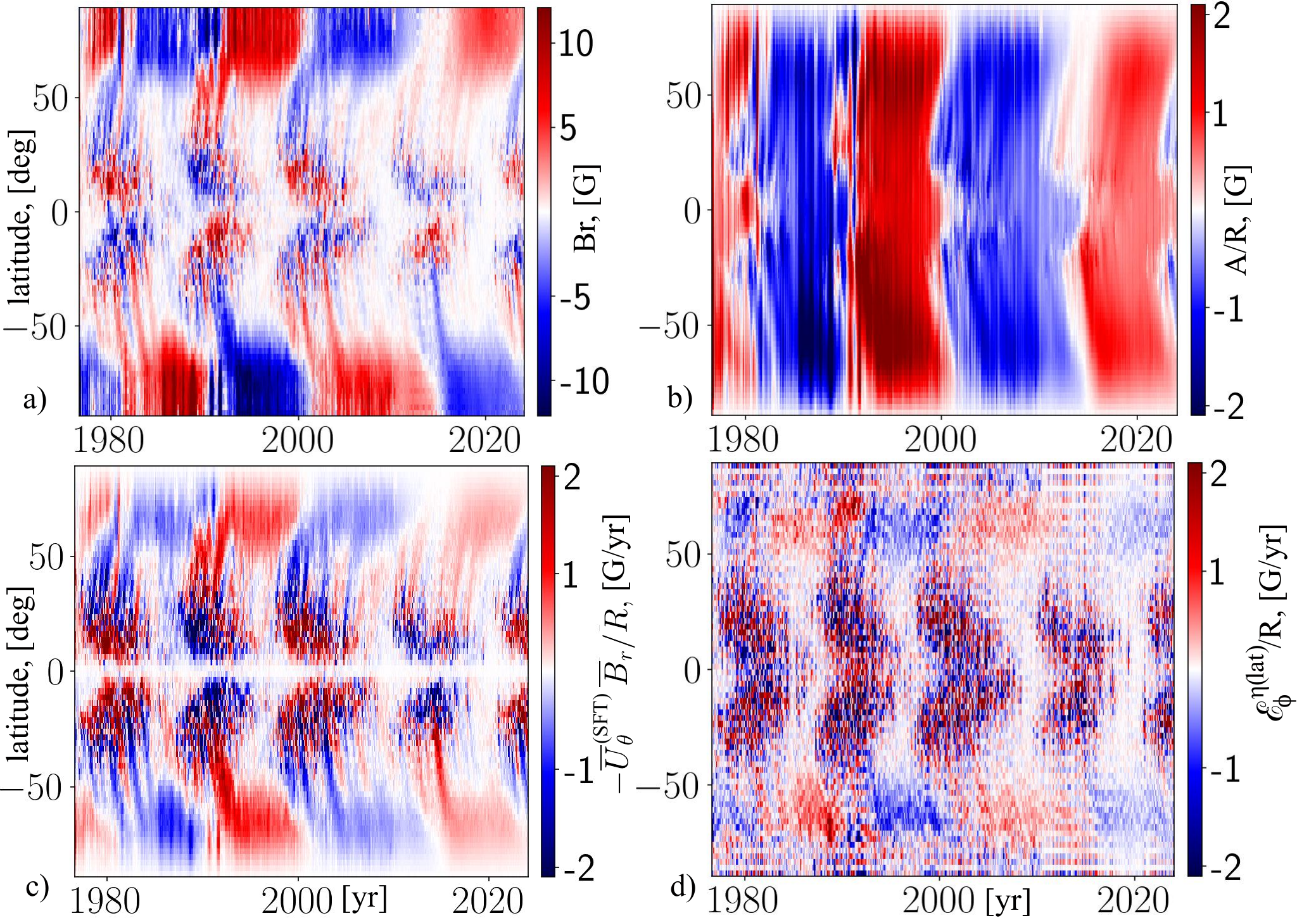}
  \caption{a) Evolution of the axisymmetric radial magnetic field of the Sun
according to data of the solar observations; b) reconstructed evolution
of the axisymmetric toroidal vector potential $A$; c) estimation
of the effect of meridional circulation in evolution $\partial_{t}A$;
d) estimation of the effect of turbulent diffusion along latitude
in evolution $\partial_{t}A$ according to observational data (see comments in the text).}\label{Fig2}
\end{figure}

\section{Observational data}

Solar observations allow us to estimate the distribution of the radial magnetic field at the photosphere. To compare the model and observations, it is convenient to reconstruct the evolution of the axisymmetric vector potential from observations. Fortunately, this is quite easy to do \citep{Blackman2003,2005AA438349K}. The method uses the solution of the equation $\overline{B}_{r}={\displaystyle \frac{1}{r\sin\theta}\frac{\partial A\sin\theta}{\partial\theta}}$ and decomposition of $\overline{B}_{r}$ and $A$ into spherical harmonics. For the magnetic field, we use synoptic maps of the radial field prepared based on observations by the Kitt Peak Observatory, the US National Solar Observatory (NSO, https://diglib-dev.nso.edu/ftp), as well as maps of the radial field with pole correction from the Solar Dynamics Observatory (SDO/HMI,  \citealp{Sun2011}). Figure \ref{Fig2}(a) shows the time-latitude diagrams of $\overline{B}_{r}$ for four solar cycles. They are obtained after averaging of the magnetic synoptic maps by azimuth. The restored toroidal vector potential $A$ is depicted in Figure \ref{Fig2}(b). The result is consistent qualitatively and quantitatively  with \citet{Blackman2003}.

{To estimate the effect of the
  meridional circulation we use the profile suggested by the surface
  flux-transport models:}
\begin{equation}
  \overline{U}_{\theta}^{(\mathrm{SFT})}=\frac{16 u_{0}}{110}\sin\left(\ensuremath{\pi-2\theta}\right)\exp\left(\ensuremath{\pi-\left|\pi-2\theta\right|}\right),\label{mobs}
  \end{equation}
{where $u_0=11$ m s$^{-1}$  \citep{Virtanen2017}. In our model the maximum magnitude of the meridional circulation is 13 m s$^{-1}$. Compared the dynamo model, the profile of Equation \ref{mobs} decreases more sharply toward the solar poles.}

{We do not take into account variations of the meridional circulation with solar cycle. The model (see  \citealp{Pipin2020}) shows that such variations can have a magnitude of 1 m s$^{-1}$. For our consideration, see Equation \ref{eq:aeq}, the cross-equatorial variations of the meridional circulation are important. In the model, such variations are small and their magnitude is less than 1 cm s$^{-1}$.}

{On the other hand, results of \cite{UptonHath2014} and \cite{Komm2022SPh} show evidence in favor of more than one order of magnitude larger meridional circulation variations at the solar equator. Results of \cite{UptonHath2014} show an amplification of dipole moment because of the time-varying meridional flow. We will return to this point later, in the discussion.}
\begin{figure}[ht]
\includegraphics[width=0.95\columnwidth]{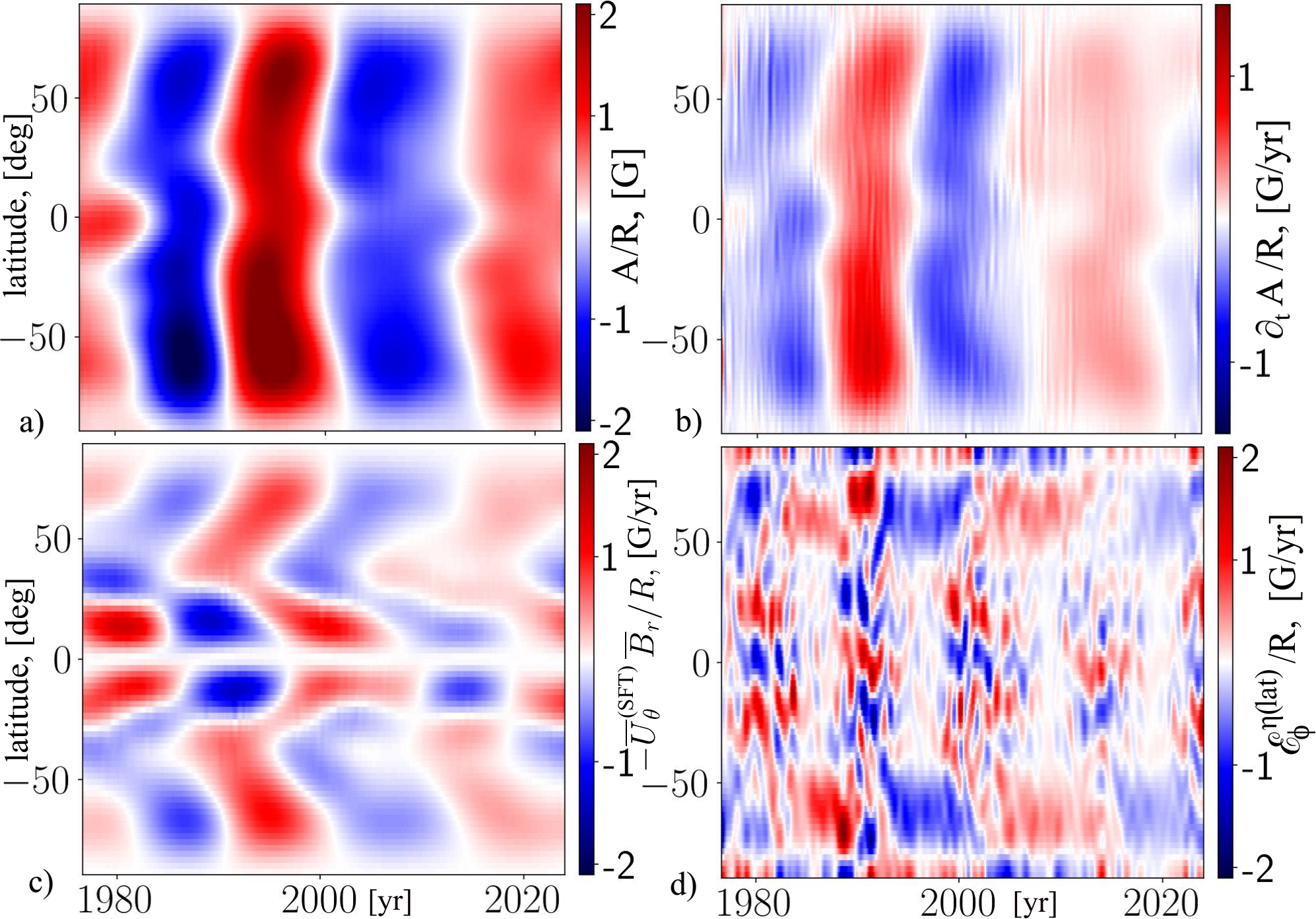} 
\caption{The same as Figure \ref{Fig2} for the time-smoothed evolution of the surface
radial magnetic field: a) the evolution of an axisymmetric toroidal
vector potential $A$; b) the derivative
of the vector potential in time, $\partial_{t}A$;  c) and d) show effects of the meridional circulation and latitudinal diffusion for the smoothed data set.}\label{Fig3}
\end{figure}

For the latitudinal turbulent diffusion, we consider an order of magnitude larger value than it is suggested by our model for the top of the convection zone; we put:
\begin{equation}
 \eta_{(\mathrm{SFT})}=12.5\eta_T,\label{etaobs}
  \end{equation}
  {where $\eta_T=5\times 10^{11}$ cm$^2$s$^{-1}$.
  While both Equations \ref{mobs} and \ref{etaobs} correspond to the choices of the “optimized” surface flux-transport models (see \citealp{Virtanen2017,Yeates2023a}), they are not  fully derived from the observations. For example, the profile of Equation \ref{mobs} does not  take into account information about the behaviour of the meridional circulation at high latitudes above 60$^{o}$, assuming that circulation speed is sharply reduced to zero toward the poles (cf. profiles of \citealp{UptonHath2014}). Also, an analysis of the sunspots decay showed that if the process is approximated by the turbulent diffusion, then the diffusion coefficient is about $10^{11}$ cm$^2$s$^{-1}$ \citep{HathChoudh2008SoPh}. The large diffusivity in the surface flux-transport models is attributed to the effect of transport of magnetic elements by  supergranules \citep{Yeates2023a}. }

\section{Results}
{Figure \ref{Fig2}(c) shows the evaluation of the effect of the meridional circulation on evolution of the vector-potential $A$. Figure \ref{Fig2}(d) does the same for the effect of turbulent diffusion on latitude. We get a strong noise in our estimation of the turbulent diffusion effect, especially at low latitudes. In polar regions, the signal is clearly visible. There, its sign is opposite to that of the effect of the meridional circulation. }
\begin{figure}[ht]
\includegraphics[width=0.95\columnwidth]{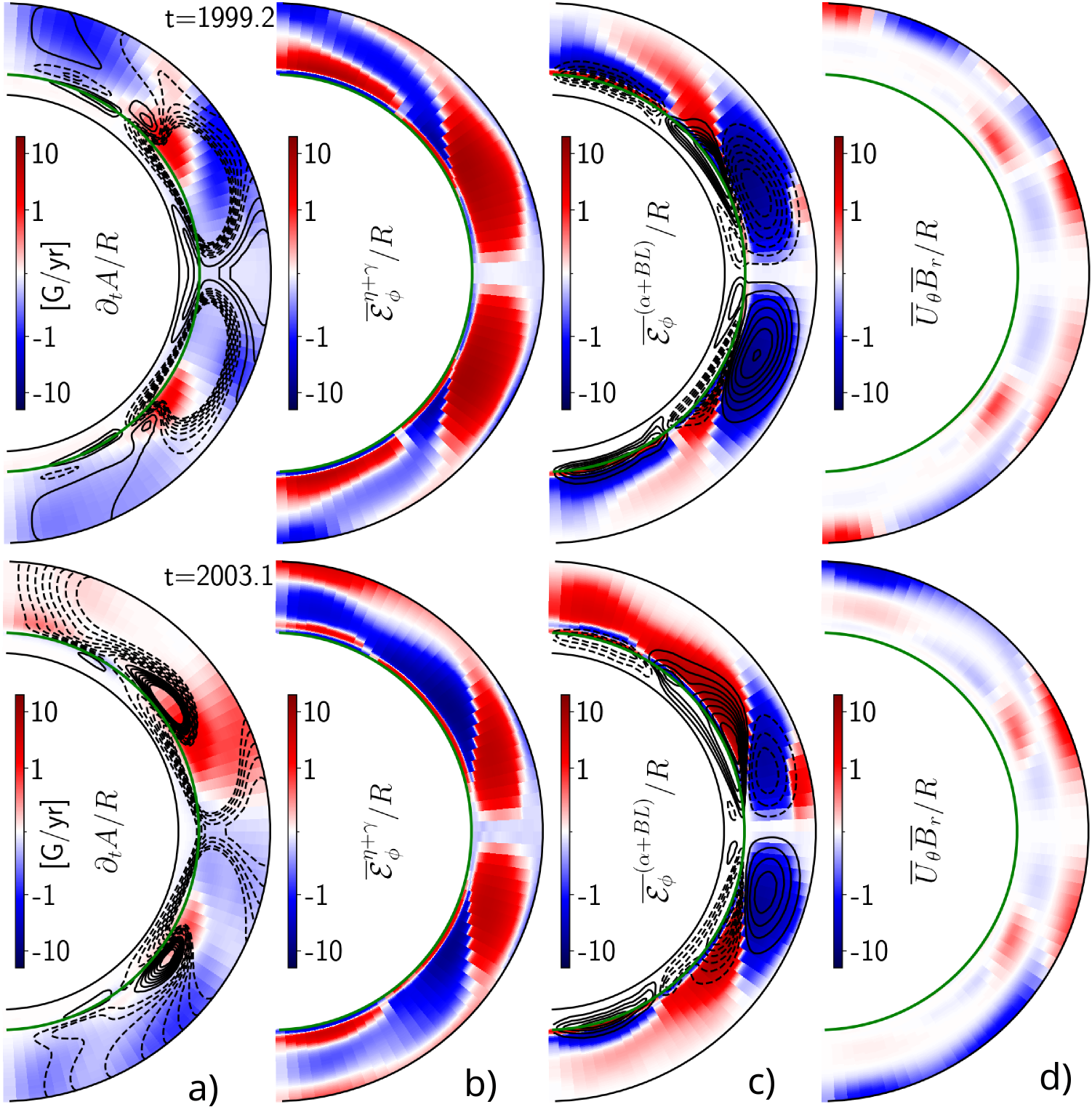}
\caption{Snapshots of distributions in the meridional section
of the convective zone for the growth and decline phase of Cycle 23:
a) streamlines of the poloidal magnetic field (dashed line -- the vector
of the field is directed counterclockwise), the amplitude of the derivative
of the vector potential is shown in color; b) the sum of the turbulent
effects of diffusion and pumping $\mathcal{E_{\phi}^{\gamma}}+\mathcal{E_{\phi}^{\eta}}$;
c) the color shows the sources of poloidal field generation, including
the Babcock--Leighton effect of BMRs, $\mathcal{E_{\phi}^{\alpha}}+\overline{\mathcal{E}_{\phi}^{\mathrm{(BMR)}}}$,
the contours show the intensity of the toroidal magnetic field fields
within range of ${\rm 3}$kG; d) the effect of meridional circulation,
$-\overline{U}_{\theta}\overline{B}_{r}$.}  \label{snap}
\end{figure}
To estimate the time derivative of the vector potential, we smoothed
the signal using Fourier filtering of harmonics with periods less
than three years. For this procedure, the standard  packages of \textsc{python scipy
} are applied. In addition, subsequent smoothing of the signal using
empirical modes is applied using \textsc{python pyemd} package. Finally,
we filter out the high spatial spherical harmonics of $\ell>13$.
The results are shown in Figure \ref{Fig3}. We see that neither the meridional
circulation nor the turbulent diffusion in latitude can explain the
time--latitude variations of the time derivative of the vector potential.
The contribution of the meridional circulation varies in antiphase
with $\partial_{t}A$. According to Equation \ref{eq:a}, the equatorial
value of the derivative $\partial_{t}A$ determines the rate of change
of the axisymmetric radial magnetic field flux through the hemisphere.
{At the solar equator, the phase of $\eta_{(\mathrm{SFT})}{\displaystyle \frac{\sin\theta}{r^{2}}\frac{\partial^{2}A}{\partial\mu^{2}}}$
seems to be following the phase of $\partial_{t}A$. For the given choice of $\eta_{(\mathrm{SFT})}$ the latitudinal turbulent diffusion can provide the required rate of change
of the hemispheric flux of the axisymmetric radial magnetic field.}
\begin{figure}[ht]
\includegraphics[width=0.7\columnwidth]{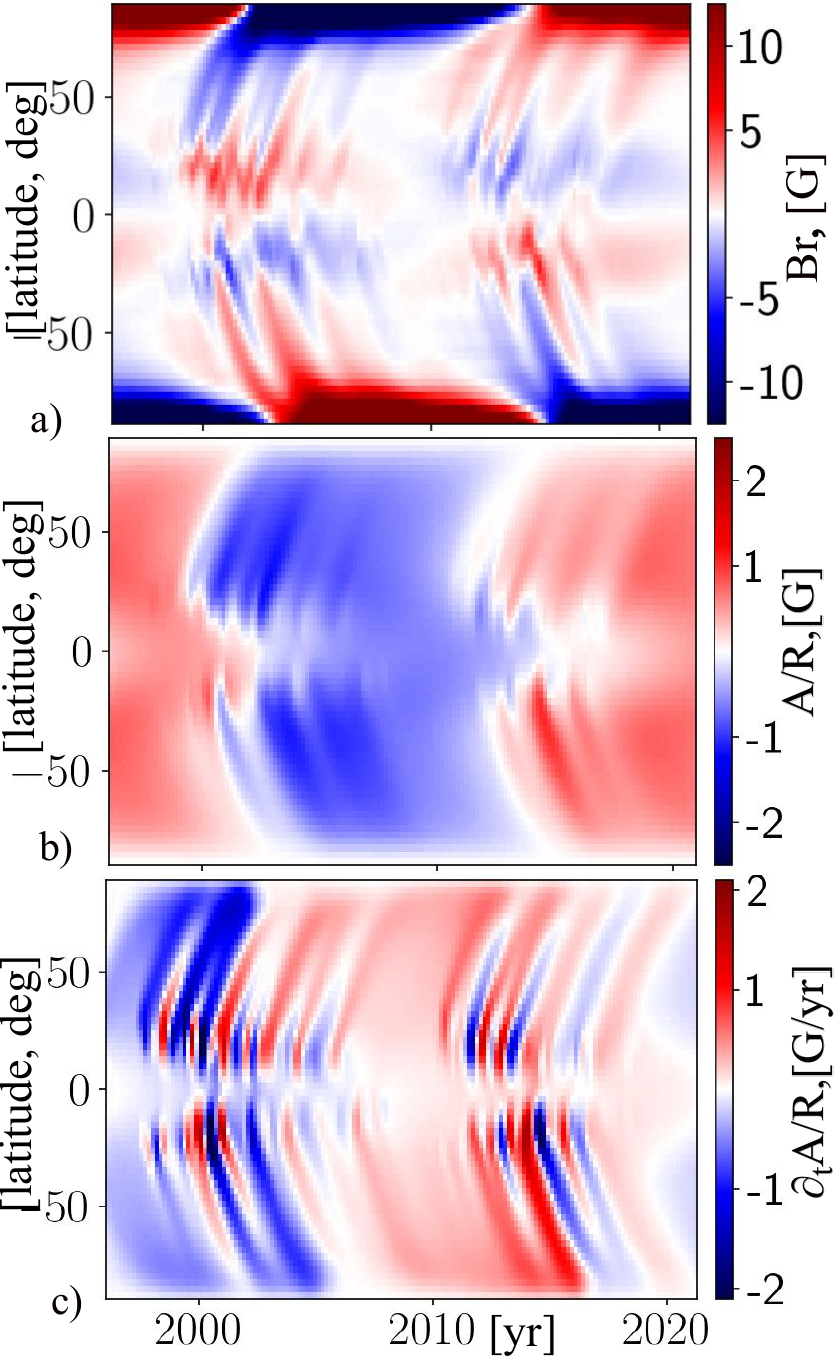}
\caption{a) The evolution of an axisymmetric radial magnetic
field at the outer boundary $r=0.99R$ in the dynamo model of the Solar
Cycle 23 and 24; b) and c) show the evolution of the vector potential
$A$ and of its derivative $\partial_{t}A$.} \label{Fig5}
\end{figure}

Let us now consider the dynamo model of Cycles 23 and 24 of \citet{PKT23}. Figure \ref{snap} shows snapshots of the distributions of the poloidal magnetic field and the turbulent effects of the dynamo evolution of the large-scale poloidal magnetic field in the meridional section
of the convective zone for the growth and decline phases of the Cycle 23. We see that the effects of diffusion and generation are in dynamic balance inside the convective zone. In the northern hemisphere, at the growth phase, the sign of the variation rate of the vector potential
in the middle and low latitudes coincides with the sign of the generation effects. At the decline phase of the magnetic cycle, when the general structure of the poloidal magnetic field is established, the sign of the variation rate of the vector potential coincides with the sign
of the turbulent diffusion effect. The maximum of the turbulent diffusion effect coincides with the center of the streamlines pattern of the poloidal magnetic field inside the convective zone.

\begin{figure}[ht]
\centering \includegraphics[width=0.65\columnwidth]{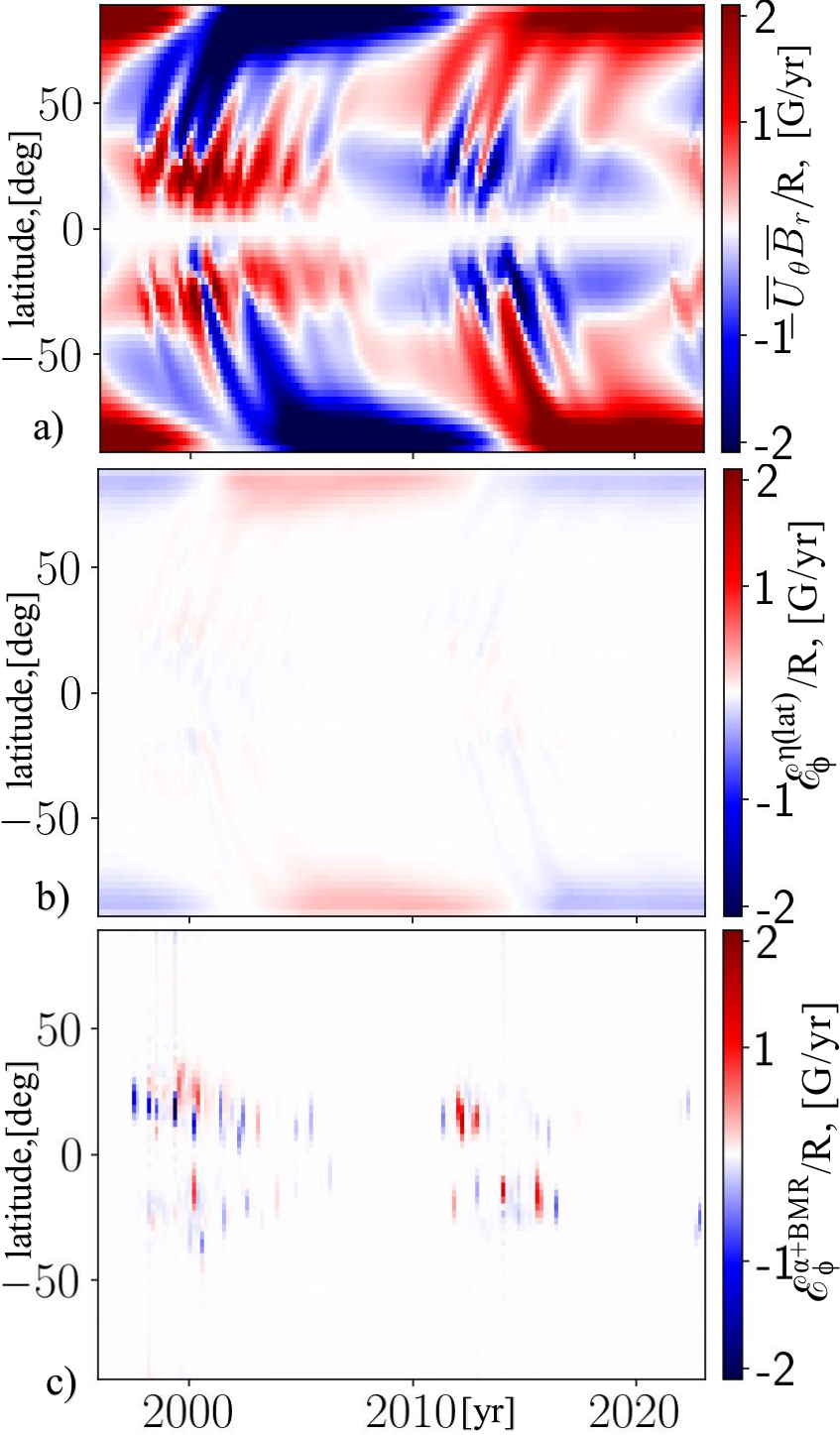}
\caption{Time--latitude diagrams of the evolution of various
contributions of the dynamo equation for $\partial_{t}A$ at the outer
boundary of the dynamo, $r=0.99R$: a) the effect of meridional circulation;
b) the effect of diffusion along the latitude, $\mathcal{E}_{\phi}^{\eta(lat)}$; c) the same as b) the dynamo generation effects, including the $\alpha$ effect and
the Babcock--Leighton effect of the BMRs.} \label{Fig6}
\end{figure}

Figures \ref{Fig5} and \ref{Fig6} show the time--latitude evolution diagrams for the radial magnetic field, vector potential, its time derivative, as well as the evolution of various contributions of turbulent electromotive force and the effect of meridional circulation. {The dynamo model evolution of the vector potential shows qualitative and quantitative agreement with diagrams of Figures \ref{Fig2}(b) and \ref{Fig3}(a). The diagram of  the parameter $\partial_t A$  allows us to identify positions of the most powerful BMRs' injections in the dynamo model, see Figure \ref{Fig5}(c). The average sign and amplitude of $\partial_t A$ agrees with the diagram of Figure \ref{Fig3}(b). Figure \ref{Fig6} shows that, in the model, as well as in the observations, the effect of meridional circulation forms an essential part of the evolution of the magnetic field on the surface.  The contribution of the latitudinal diffusion effect to the signal of $\partial_t A$ is relatively small compared to the diagram of Figure \ref{Fig3}(d).} The local contribution of the effects of generating a large-scale poloidal magnetic field because of emergence of bipolar regions on the surface is significantly less in amplitude than the local values of the signal $\partial_{t}A$. 

\begin{figure}
\centering \includegraphics[width=0.65\columnwidth]{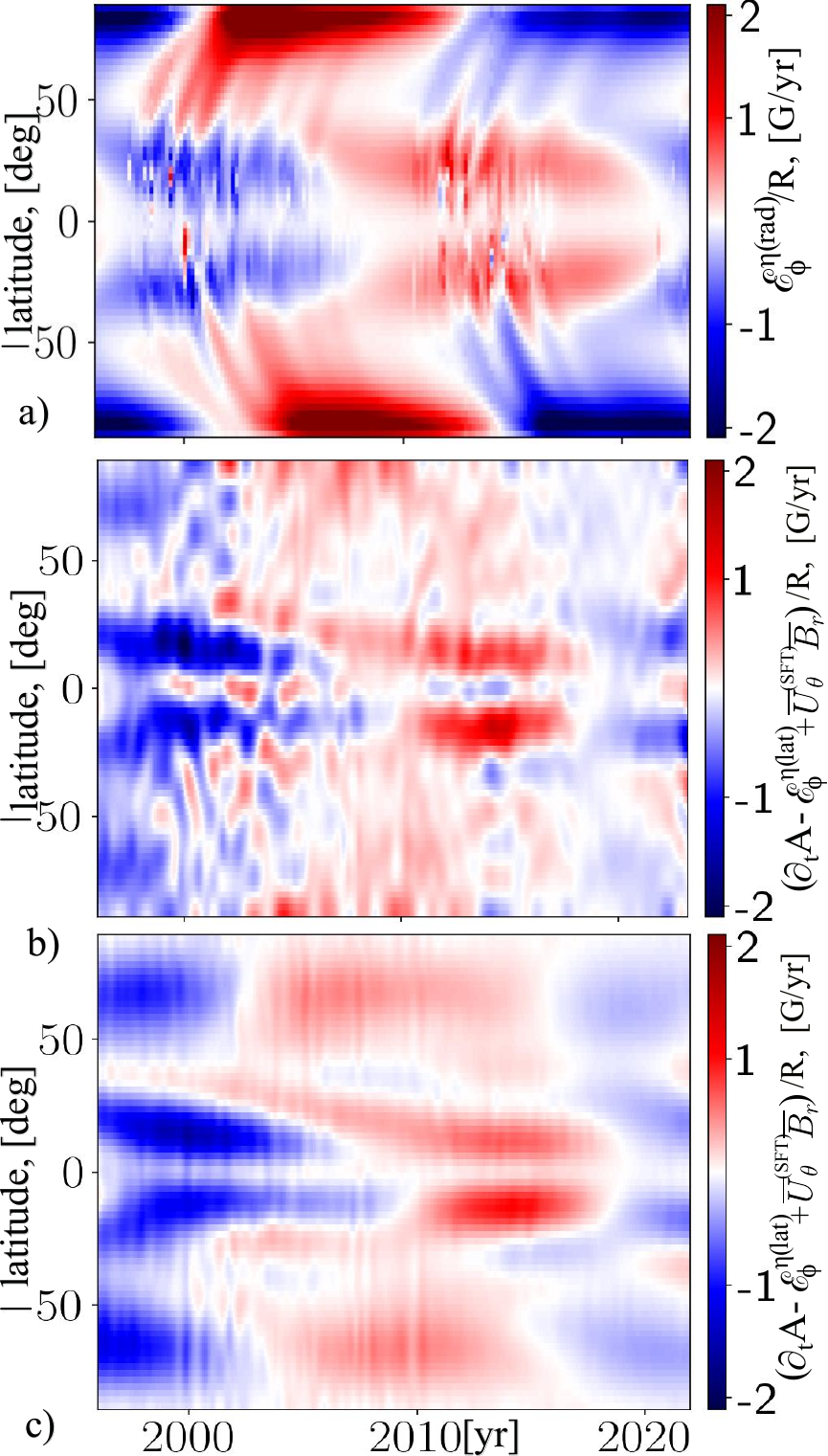}
\caption{(a) The time--latitude diagram for the effect of the radial turbulent diffusion of a poloidal magnetic field, $\mathcal{E}_{\phi}^{\eta(rad)}$ at the top boundary $r=0.99R$; (b) the result of calculation of  $\partial_{t}A+\overline{U}^{(\mathrm{SFT})}_{\theta}\overline{B}_r-\mathcal{E}^{\eta(lat)}_{\phi}$ using the smoothed version of the data set from observations, see Figures \ref{Fig3}(b), \ref{Fig3}(c) and \ref{Fig3}(d); (c) shows the same as (b) but for the order of magnitude smaller coefficient of the turbulent diffusion, i.e., the same as in the dynamo model.}\label{Fig6b}
\end{figure}

{Figure \ref{Fig6b}(a) shows the time--latitude diagram  of the radial turbulent diffusion effect on the surface vector-potential in the dynamo model. We find that the sum of the effects of the radial turbulent diffusion and the transport effects by the meridional circulation gives the time-latitude variations of the signal $\partial_{t}A$ (see Figure \ref{Fig5}(c)).
Now, let us subtract the sum of diagrams in Figures \ref{Fig3}(c) and \ref{Fig3}(d) from the diagram of Figure \ref{Fig3}(b). In other words, we calculate $\partial_{t}A+\overline{U}^{(\mathrm{SFT})}_{\theta}\overline{B}_r-\mathcal{E}^{\eta(lat)}_{\phi}$ 
using the smoothed version of the data set from observations. Figure \ref{Fig6b}(b) shows the result of calculation for the turbulent diffusion coefficient $\eta_{(SFT)}$, see Equation \ref{etaobs}. At low latitudes, the diagrams  of Figures \ref{Fig6b}(a) and \ref{Fig6b}(b) agree qualitatively. The result of calculation in Figure \ref{Fig6b}(c) corresponds to the turbulent diffusion coefficient of the dynamo model. This diagram shows a better agreement with Figure \ref{Fig6b}(a) than that in Figure \ref{Fig6b}(b). There is a phase difference of  Figures \ref{Fig6b}(a) and \ref{Fig6b}(c) in polar regions. It is likely because of the different behavior of the meridional circulation profiles of the dynamo model and the profile of Equation \ref{mobs}. Nevertheless, these findings may trace for the effect of the radial turbulent diffusion in evolution of the surface axisymmetric radial magnetic field.}

\begin{figure}[ht]
\includegraphics[width=0.95\columnwidth]{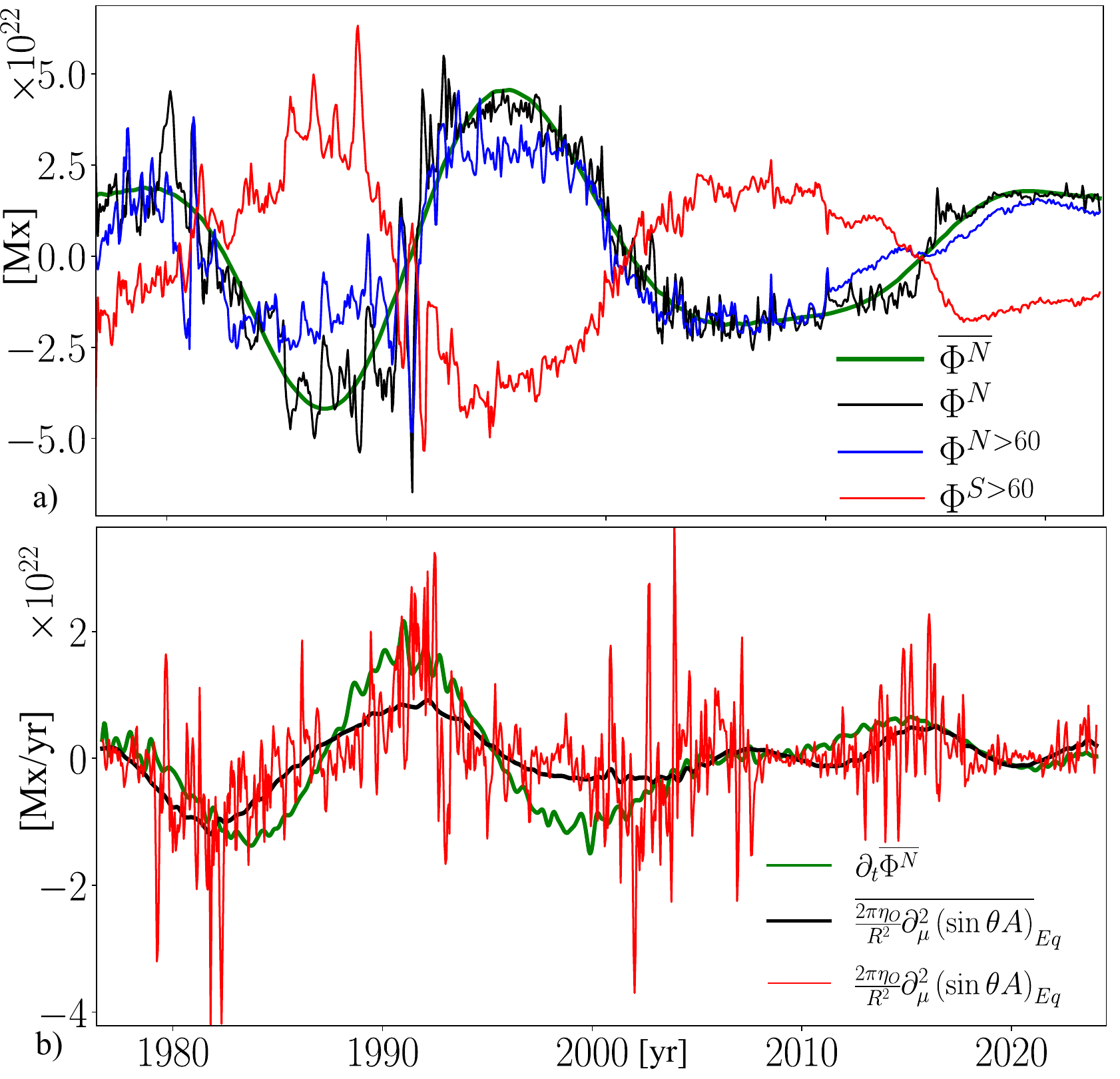}
\caption{(a) The flux of an axisymmetric radial magnetic field through
the surface of the photosphere in the northern hemisphere, $\Phi^{N}$,
($\overline{\Phi^{N}}$ is the time-smoothed version of the signal),
$\Phi^{N>60}$ and $\Phi^{S>60}$ are the fluxes of the radial magnetic
field through the polar caps; (b) $\partial_{t}\overline{\Phi^{N}}$
-- the rate of change of the radial magnetic field flux in the northern
hemisphere, while $\frac{2\pi\eta_{T}}{R}\partial_{\mu}^{2}A$ is the contribution
of turbulent diffusion along the latitude to the derivative of the
magnetic field flux, a similar value in the horizontal bar shows the
time-smoothed signal.} \label{obs}
\end{figure}

{Figure \ref{obs}a shows the evolution of the hemispheric flux of the axisymmetric radial magnetic field, $\overline{\Phi}^{N}$, through the surface of the photosphere in the northern hemisphere over Solar Cycles 21--24 . Interestingly, the hemispheric flux of the radial magnetic field in Cycles 21 and 22 differs considerably from the flux of the magnetic field through the polar caps (latitude $\ge60^{\circ}$).  The rate of generation of the hemispheric flux, $\partial_{t}\overline{\Phi}^{N}$, varies with the phase shift of $\pi/2$ relative to $\overline{\Phi}^{N}$. As expected, according to the general dynamo scenario, the maximum values of $\partial_{t}\overline{\Phi}^{N}$ correspond in time to the maximum of the sunspot cycle. }

{Our estimates confirm that the magnitude $\eta_{\mathrm(SFT)}=5\cdot 10^{12}$ cm$^2$s$^{-1}$ is enough to reproduce the generation rate $\partial_t\overline{\Phi}^{N}$. Nevertheless, the signal of the turbulent diffusion effect is, sometimes out of the phase of $\partial_t\overline{\Phi}^{N}$. For example,  it happens at the growing phase of the Solar Cycle 23.}

\begin{figure}[ht]
\includegraphics[width=0.99\columnwidth]{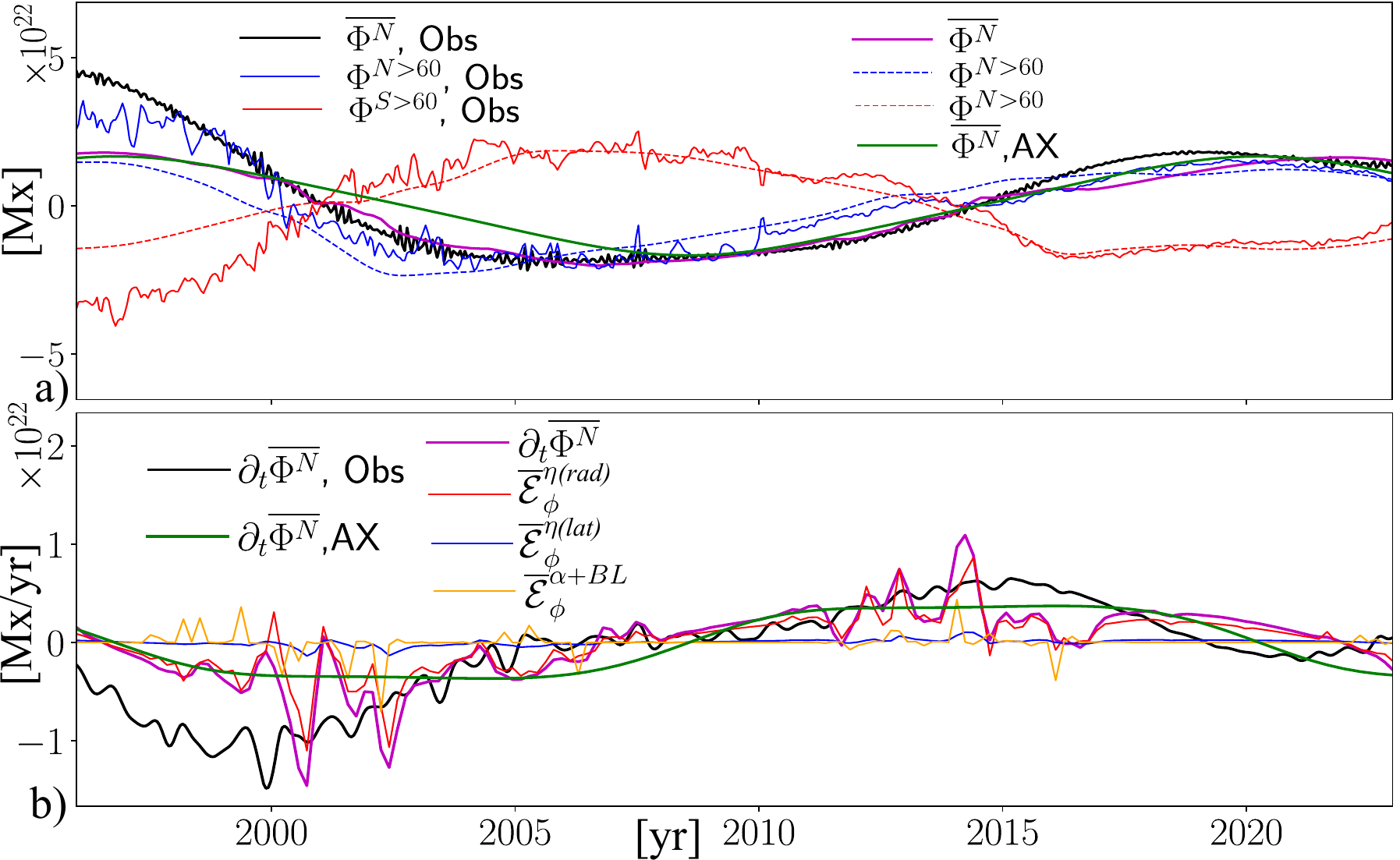}
\caption{The same as Figure \ref{obs} for the dynamo models. Panel (b) additionally
shows the contributions of the surface effects of turbulent diffusion
and generation to the rate of change of the radial magnetic field
flux in the northern hemisphere. The green lines in both panels show results for the axisymmetric dynamo model of \cite{PKT23}.} \label{fig:mod}
\end{figure}

{A comparison with the results of the model (see Figure \ref{fig:mod}) shows that the model reproduces the evolution of $\overline{\Phi}^{N}$, as well as the fluxes of the radial magnetic field through the polar caps. 
From the point of view of the dynamo model, the most significant contribution to the $\overline{\Phi}^{N}$ signal is connected to the effect of radial turbulent diffusion. For comparison, we show the results for the hemispheric flux of the radial magnetic field and its generation rate in the axisymmetric dynamo model of \cite{PKT23}. We see that the large portion of the dynamo generated flux of the surface large-scale magnetic field is because of the axisymmetric dynamo operating in the bulk of the convection zone.}

\section{Discussion and conclusions}

In this study, we investigated the rate of generation of a radial magnetic field flux through a hemisphere in a solar dynamo using observational data  and a mean-field dynamo model of \cite{PKT23}. The model takes into account the turbulent generation of a poloidal magnetic field associated with the alpha effect and the Babcock--Leighton effect, which results from the emergence of tilted bipolar sunspots groups. We process the observational data set using the parameters of the surface flux-transport models.

Both the dynamo model and observations indicate  that the hemispheric flux generation rate can depends strongly on the effect of turbulent diffusion. Interesting that the surface effects because of either turbulent generation by the alpha effect or because of the Babcock--Leighton effect are not so profound. In both cases, the model describe the generation process by effective turbulent electromotive force (EMF),  $\mathbf{\mathcal{E}_{\phi}^{\alpha}+\overline{\mathcal{E}_{\phi}^{(BMR)}}}$, along the axisymmetric toroidal magnetic field $\overline{B_{\phi}}$ \citep{Krause1980}. 

{Such an EMF generates the poloidal magnetic field. The streamlines of this magnetic field go around the axis of a toroidal magnetic field. The polarities of the field are opposite from the south and north of the $\overline{B}_{\phi}$ axis, and also in the radial direction, respectively. The efficiency of the radial magnetic flux production depends on the boundary conditions and the turbulent diffusion along the radius and latitude.} If we consider the evolution of the poloidal magnetic field along latitude, then the amplitude of the generation of $\overline{\Phi}^{N}$ in the northern hemisphere depends on how the southern polarity of the poloidal field generated by  $\mathbf{\mathcal{E}_{\phi}^{\alpha}+}\overline{\mathcal{E}_{\phi}^{\mathrm{(BMR)}}}$ and cancels with the corresponding northern polarity from the southern hemisphere. This determines the efficiency of the Babcock--Leighton scenario and models of surface magnetic flux transport (surface flux transport (SFT),  \citealp{Yeates2023a}). The authors of these models recognize the need to take into account radial diffusion (see, e.g., \citealp{Virtanen2017} and the review cited above). {Our results suggest that the required hemispheric generation rate could be reconciled in the SFT models with an order of magnitude less turbulent diffusion if in the radial direction is taken into account.}

In a solar dynamo, the variations of the large-scale magnetic field is much smaller in radius than in latitude. This leads to a relatively high role of turbulent diffusion along the  radius in the presented mean-field dynamo model. Our dynamo model operates in weakly  nonlinear regime with the $\alpha$ effect, which is slightly above (about 10\%) the dynamo threshold. For such a situation, the additional Babcock--Leighton effect of surface BMRs activity shows a profound contribution, and it leads to an increase in the dynamo efficiency. The tilted loops of the BMRs increase the radial field flux penetrating from the depth of the convective zone by generating new flux outside the dynamo region and by reconnection of magnetic fields at the base of BMRs with an axisymmetric poloidal magnetic field inside the convective zone, because of the effects of radial diffusion. In addition, the radial turbulent diffusion maintains the poloidal field structure covering the entire convective zone (see Figure \ref{snap}). Here, the generation of a toroidal magnetic field occurs throughout the entire volume of the convective zone not only because of differential rotation in latitude (cf., the BL scenario of \citealt{CS2023SSRv}), but also radial gradients of angular velocity near the boundaries of the convection zone. Similar conclusions were drawn recently from an analysis of the toroidal magnetic flux budget using the same dynamo model by \cite{PK24} and the 3D magnetohydrodynamic simulations by \cite{FinleyStrug2024}. 

{In this study we do not consider the effects of the solar-cycle variations of the meridional circulation. Following the results of \cite{UptonHath2014} existence of such variations can greatly affect our conclusions. 
The circulation speed affects the differential rotation of the Sun \citep{Hazra2023a}. Therefore, in comparing the dynamo model and observation, the  desired picture  of the meridional circulation variations should be physically consistent with the mass and angular momentum conservation of the large-scale flow in the bulk of the convection zone. Also, the knowledge  about the subsurface magnetic field evolution is needed, in order to distinguish between the effects of the meridional circulation variations and turbulent diffusion.}

\begin{acknowledgements}
The author thanks the Ministry of Science and Higher Education of
the Russian Federation for financial support (Subsidy No.075-GZ/C3569/278).
I thank the anonymous referee for constructive comments and suggestions.
\end{acknowledgements}
 
\begin{authorcontribution}
The author takes full responsibility for results  presented in the paper.
\end{authorcontribution}

\begin{fundinginformation}
The author thanks the Ministry of Science and Higher Education of
the Russian Federation for financial support (Subsidy No.075-GZ/C3569/278)
\end{fundinginformation}

\begin{dataavailability}
The data of the dynamo simulations are available from the author  by
request.
Observational data credit to the USA National
Solar Observatory (NSO, https://diglib-dev.nso.edu/ftp ), as well
as SDO/HMI observations and SDO/HMI pipeline for synoptic maps of the radial field with pole correction \citep{Sun2011}
\end{dataavailability}

\begin{materialsavailability}
All additional materials related to the paper can be available from
the author by request.
\end{materialsavailability}

\begin{codeavailability}
The dynamo code is available by request from the author.
\end{codeavailability}

\begin{ethics}
\begin{conflict}
The author declares that he has no conflicts of interest.
\end{conflict}
\end{ethics}
\bibliographystyle{spr-mp-sola}

\end{document}